\documentclass[10pt]{article}
\usepackage[OE]{express}

\begin{document}
\title{Dual-phase-modulated plug-and-play measurement-device-independent continuous-variable quantum key distribution}

\author{Qin Liao,\authormark{1} Ying Guo,\authormark{1}, Yijun Wang,\authormark{1}, Duan Huang,\authormark{1}}

\address{\authormark{1}School of Information Science and Engineering, Central South University, Changsha 410083, China\\
State Key Laboratory of Advanced Optical Communication Systems and Networks, \\Shanghai Jiao Tong University, Shanghai 200240, China 
}

\email{\authormark{*}yingguo@csu.edu.cn} 



\begin{abstract}
We suggest an improved plug-and-play measurement-device-independent (MDI) continuous-variable quantum key distribution (CVQKD) via the dual-phase modulation (DPM), aiming to solve an implementation problem with no extra performance penalty. The synchronous loophole of different lasers from Alice and Bob can be elegantly eliminated in the plug-and-play configuration, which gives birth to the convenient implementation when comparing to the Gaussian-modulated coherent-state protocol. While  the local oscillator (LO) can be locally generated by the trusted part Charlie, the LO-aimed attacks can be accurately detected in the data post-processing. We derive the security bounds of the DPM-based MDI-CVQKD against optimal Gaussian collective attacks. Taking the finite-size effect into account, the secret key rate can be increased due to the fact that almost all raw keys of the MDI-CVQKD system can be fully exploited for the final secret key generation without sacrificing raw keys in parameter estimation. Moreover, we give an experimental concept of the proposed scheme which can be deemed guideline for final implementation.
\end{abstract}



\section{Introduction}

Quantum key distribution (QKD) \cite{BB84, ref2} is a branch of quantum cryptography, whose goal is to provide an elegant way to allow two remote legitimate partners (Alice and Bob) to generate a random secure key with unconditional security \cite{ref3,ref4} over insecure quantum and classical channels.
There are two approaches to implement QKD, i.e., discrete-variable (DV) QKD \cite{ref5,ref6} and continuous-variable (CV) QKD \cite{ref7,ref8,ref9,ref10}. In DVQKD, the polarization states of a single photon are usually exploited to transmit the information of key bits, whereas in CVQKD, the sender, Alice, usually encodes key bits in the quadratures ($\hat{x}$ and $\hat{p}$) of the optical field with Gaussian modulation \cite{ref11}, and the receiver, Bob, can restore the secret key bits through high-speed and high-efficiency homodyne or heterodyne detection techniques \cite{ref12}. 

Currently, the CVQKD protocol has been implemented through the established standard telecommunication facilities, which is more convenient and practical than its DVQKD counterpart. However, there is usually an assumption that devices are perfect and cannot be eavesdropped by the third untrusted party, and consequently doubts about the immaculate CVQKD system have been raised. An ideal model of CVQKD is not enough for the security  analysis of the practical CVQKD system. For example, there is no need to consider the effect of local oscillator (LO) in ideal models but it is necessary to take it into account in the practical CVQKD system, since eavesdroppers may exploit the transmitted LO to launch practical attacks such as wavelength attacks \cite{ref14,ref15}, saturation attacks \cite{ref16}, calibration attacks \cite{ref17}, and LO fluctuation attacks \cite{ref18}. Furthermore, the imperfections of detectors can be maliciously exploited, which make the CVQKD system vulnerable to various attacks. To remove all existing and yet-to-be-discovered detector side channels, measurement-device-independent (MDI) QKD was proposed \cite{ref188,ref19,ref20}. It offers an immense security advantage over standard security proofs and has the power to double the secure distance \cite{ref19}. 

So far, much progress has been made in MDI-DVQKD \cite{ref21,ref22,ref23,ref24,ref25} and MDI-CVQKD \cite{ref26,ref27,ref277,ref28}. In a MDI-CVQKD protocol, both Alice and Bob are senders while an untrusted third party Charlie is introduced to realize Bell measurements. Such measurement results will be used by Alice and Bob in post-processing to generate the secure keys. MDI-CVQKD has become an important research limelight since it has many practical advantages, especially for a metropolitan QKD network \cite{ref29}. Subsequently, C. Lupo et al. presented the composable security proof of MDI-CVQKD against coherent attacks \cite{ref30}. However, the theoretical feasibility does not usually mean its experimental implement although its theoretical security has been proven. In the other words, the implementation of MDI-CVQKD at present may be impractical. For example, a strong light (LO) with weak signal light is required for realizing the practical CVQKD \cite{ref31}. Because these two lights have to be precisely interfered in homodyne or heterodyne detector,  the synchronization of the two beams is crucial to implement CVQKD communication. While this problem magnifies doubly in MDI-CVQKD with two senders Alice and Bob, it renders MDI-CVQKD hard to implement stably. Moreover, the Gaussian-modulated coherent states are prepared for Alice and Bob with symmetrical modulation, which is usually implemented by applying an amplify modulator (AM) and a phase modulator (PM). Unfortunately, most of the widely used AMs, e.g. LiNbO$_3$ modulators, are polarization sensitive and features a polarizer, which means the part of light cannot be transmitted if its orientation is not aligned correctly\cite{ref32}. As a result, the performance of the practical MDI-CVQKD system will be degenerated.

To solve the above-mentioned problems, in this paper, we suggest a plug-and-play (PP) scheme for MDI-CVQKD via the dual-phase modulation (DPM). In particular, the proposed scheme waives the necessity of propagation of the LO through the insecure quantum channels from Alice's and Bob's sides. Instead, the real LO is generated from the same laser of quantum signal at Charlie's side, and thus it avoids the problems of synchronization of different lasers as well as the LO-aimed attacks. Moreover, the reference of two signals can be guaranteed identically and the polarization drifts can be compensated automatically since only one laser is needed for the PP DPM-based MDI-CVQKD. Meanwhile, a polarization-insensitive dual-phase modulation strategy is adopted to Alice's and Bob's sides respectively which shows the experimental feasibility of preparing Gaussian states. We  derive the security bounds against optimal Gaussian collective attacks, which shows that the proposed scheme works equivalently to symmetrically modulated Gaussian-state MDI-CVQKD protocols. When taking into account the finite-size effect almost all raw keys generated by the proposed scheme can be used for final secret key generation, without sacrificing any raw keys in parameter estimation. 

This paper is structured as follows. In Sec. II, we describe the traditional MDI-CVQKD protocol, and subsequently propose the PP DPM-based  MDI-CVQKD. In Sec. III, we focus on the security analysis with numerical simulation in asymptotic limit and finite-size regime. Finally, conclusions are drawn in Sec. IV.

\section{The PP DPM-based MDI-CVQKD protocol}

To make the derivation of the PP DPM-based MDI-CVQKD protocol self-contained, we illustrate the characteristics of the MDI-CVQKD protocol and then extend it to the PP DPM-based MDI-CVQKD protocol.

\subsection{Characteristics of the MDI-CVQKD protocol}
\begin{figure}[htbp]
\centering\includegraphics[width=13cm]{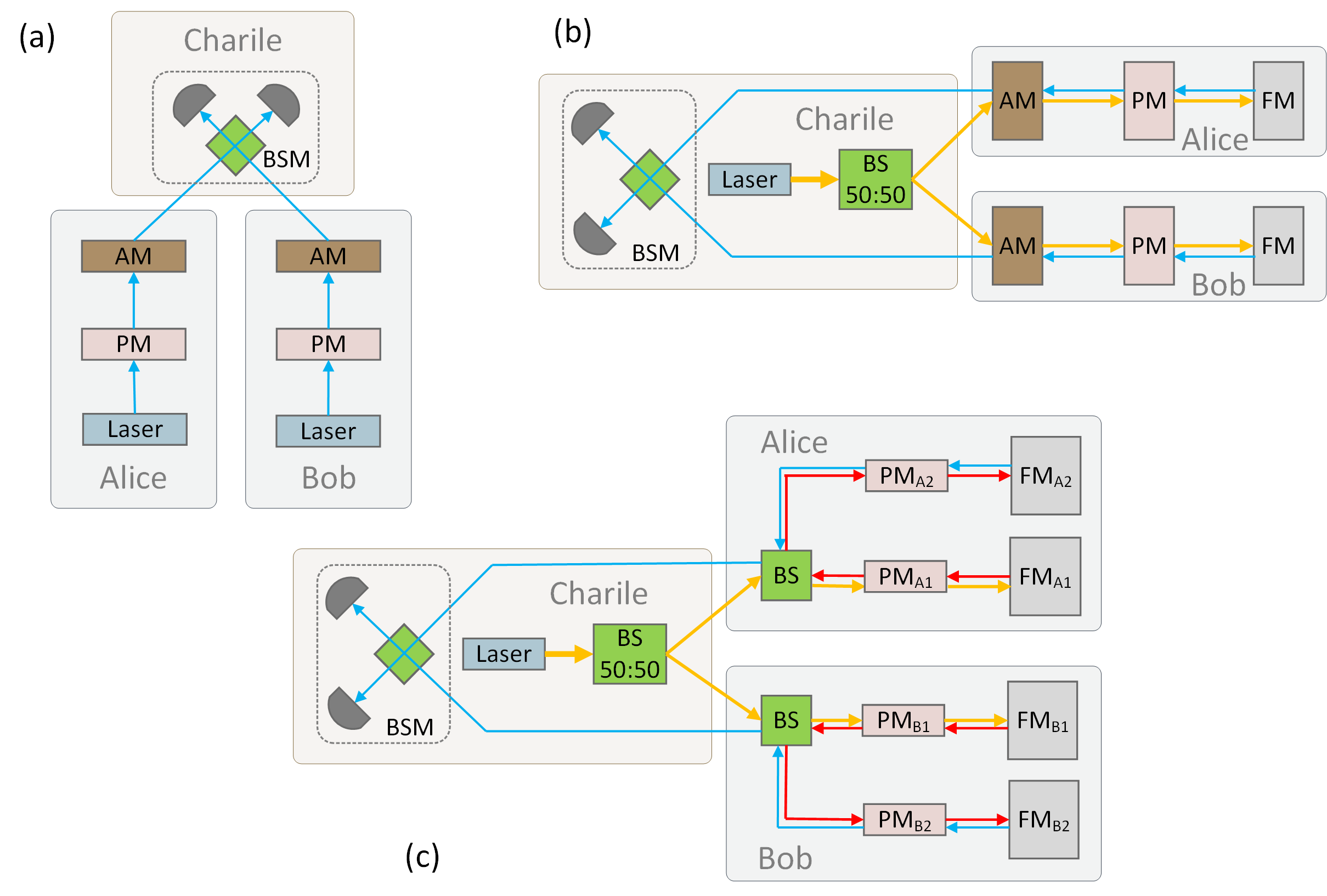}
\caption{Schematic diagrams of (a) traditional MDI-CVQKD protocol. Alice and Bob prepare coherent states independently, and send them to Charlie for Bell state measurement. (b) PP   MDI-CVQKD protocol. Charlie initially launches pulses to Alice and Bob, and then Alice and Bob reflect back the pulses to Charlie after Gaussian modulation. (c) PP DPM-based MDI-CVQKD protocol. Charlie still initially launches pulses to Alice and Bob, Alice and Bob respectively use dual-phase-modulation strategy to encode the information and subsequently send the pulses back to Charlie. AM, Amplitude modulator; PM, Phase modulator; FM, Faraday mirror; BS, Beam splitter; BSM, Bell state measurement.}
\label{MDI}
\end{figure}

In the MDI-CVQKD protocol, the side-channel attacks can be eliminated since one does not need to make any assumption on the measurement device. As shown in Fig. \ref{MDI}.(a), two lasers are adopted to Alice's and Bob's side respectively, where each side modulates information independently using AM and PM. After that, the two Gaussian-modulated pulses are sent to Charlie who measures the incoming modes with Bell state measurement (BSM). In particular, the prepare-and-measure model of MDI-CVQKD protocol can be described as follows.

$Step \; 1. $
Alice randomly prepares a coherent state with complex amplitudes $\alpha'=(x_A'+ip_A')/2$ and Bob randomly prepares another coherent state with complex amplitudes $\beta'=(x_B'+ip_B')/2$, where the local variables $X'=(x_A', p_A')$ and $Y'=(x_B', p_B')$ are Gaussian distributed with variances $V_A$ and $V_B$, respectively. Then Alice and Bob send their coherent states to Charlie.

$Step \; 2. $
After receiving the transmitted coherent states, Charlie performs BSM-based detections. The coherent states are interfered and measured by the homodyne detectors, and the measurement results described as variable Z with complex value $\gamma=(x_Z+ip_Z)/2$ are announced by Charlie.

$Step \; 3. $
When Alice and Bob receive Charlie's measurement results, they can estimate the covariance matrix $\Gamma_{XYZ}$ of the tripartite state $\rho_{XYZ}$, which  can be used for the security analysis \cite{ref11,ref33,ref34}.

$Step \; 4. $
Alice and Bob modify their data as $X=(x_A, p_A)$ and $Y=(x_B, p_B)$, where
\begin{equation}
\begin{aligned}
x_A&=x_A'-k_{x_A'}(\gamma), \; p_A=p_A'-k_{p_A'}(\gamma), \\ 
x_B&=x_B'-k_{x_B'}(\gamma), \; p_B=p_B'-k_{p_B'}(\gamma).
\end{aligned}
\end{equation}
Here $k$ is the amplification coefficient related to channel loss, and the variables $X$ and $Y$ represent the local raw keys of Alice and Bob, respectively.

$Step \; 5. $
Alice and Bob use an authenticated public channel to finish the error correction and privacy amplification, and finally generate the identical secret key.

\begin{figure}[htbp]
\centering\includegraphics[width=13cm]{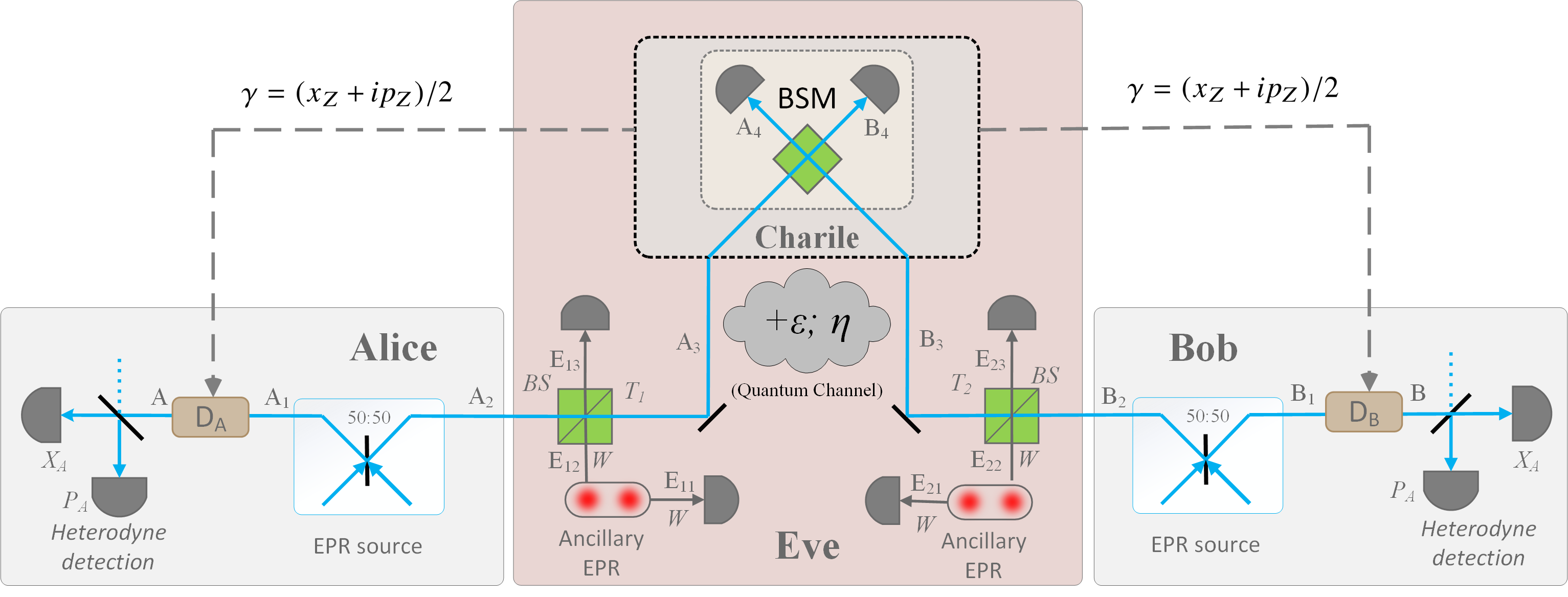}
\caption{Entanglement-based model of MDI-CVQKD protocol with entangling cloner attacks. Alice and Bob respectively generate EPR pairs and send them to the third party Charlie through the untrusted quantum channel. Charlie measures the incoming modes using BSM and subsequently sends the measurement result back to Alice and Bob. $\varepsilon$ is excess noise and $\eta$ is the transmittance of the quantum channel.}
\label{MDIsec}
\end{figure}

The prepare-and-measure model is usually easy to apply, while considering its security, the prepare-and-measure model is equivalent to the corresponding entanglement-based model for convenient security analysis \cite{ref35}. As shown in Fig. \ref{MDIsec}, Alice and Bob first prepare two-mode squeezed vacuum state [Einstein-Podolsky-Rosen (EPR) state] respectively. Each sender keeps one mode $A_1(or B_1)$ and sends another mode $A_2(or B_2)$ to the third party (Charlie) through the untrusted quantum channel. An eavesdropper, says Eve, may replace the quantum channel between the two senders and Charlie with her own quantum channel to launch the entangling cloner attacks, which has been proven to be one kind of the optimal Gaussian collective attack \cite{ref36,ref37}.
The incoming modes $A_3$ and $B_3$ are received by Charlie and subsequently interfered at a beam spiltter (BS) with two output modes $A_4$ and $B_4$. Then both the $x$-quadrature of $A_4$ and the $p$-quadrature of $B_4$ are measured by homodyne detectors, and the measurement result $\gamma=(x_Z+ip_Z)/2$ are announced by Charlie. After receiving Charlie's measurement results, Alice and Bob respectively displace their own mode $A_1(or B_1)$ by operations $\hat{D}_A(\gamma)$ and $\hat{D}_B(\gamma)$. Finally, the yielded modes $A$ and $B$ are measured by heterodyne detectors to generate the raw key $\{X,Y\}$.

\subsection{Design of the PP MDI-CVQKD protocol}

The plug-and-play (PP) MDI-CVQKD protocol can be designed on the basis of the traditional MDI-CVQKD protocol by putting the laser at Charlie's side instead of applying two lasers at Alice's and Bob's sides. 
As shown in Fig. \ref{MDI}(b), the data-processing is almost similar, but slightly different from that of the traditional MDI-CVQKD protocol. Firstly, Charlie sends strong coherent light to Alice and Bob through a 50:50 beam splitter. Each splited light is transmitted through an optical fiber, and then reflected by a faraday mirror (FM) of Alice (or Bob). An AM and a PM are adopted at each side to encode information using Gaussian modulation. Then, the light is sent back to Charlie, and then BSM and data post-processing are followed to generate the secret keys.

The PP MDI-CVQKD protocol has several remarkable features. It is similar to the traditional MDI-CVQKD because the signal sent from Charlie to Alice and Bob does not contain any Gaussian-modulated information, and hence the useful information is only available in the trusted parts. In addition, the mode matching issue such as the problem of synchronization can be solved since the lights of Alice and Bob are generated from the same laser so that the spectral modes of the pulses are identical. From the viewpoint of its experimental implementation, the polarization drift during the optical fiber transmission can be automatically compensated. 

It is necessary to show the effect of the PP configuration on the security of the realistic MDI-CVQKD. The security against any detector side channel attacks is guaranteed since the PP configuration does not disturb the measurement setup of the MDI-CVQKD. Moreover, one does not need to transmit LO through the untrusted channel, but can generate LO locally at Charlie's side, which eliminates all LO-aimed attacks in the security analysis.

\subsection{Design of the PP DPM-based MDI-CVQKD protocol}

So far, the Gaussian-modulated CVQKD protocols, including the Gaussian quantum state, Gaussian operation and Gaussian measurement, are experimentally feasible and simple for the mathematical description \cite{ref29}. As a result, the traditional CVQKD protocols are usually based on Gaussian modulation except some discretely-modulated CVQKD protocols \cite{ref39,ref40,ref41}. Theoretically, the Gaussian quantum state can be prepared by using AM and PM. Unfortunately, most of the widely used AMs, e.g. LiNbO$_3$ modulators, are polarization sensitive and features a polarizer, where the part of light cannot be transmitted if its orientation is not aligned correctly\cite{ref32}, resulting in the  degenerated performance in practice.
To solve this problem, we suggest the PP DPM-based MDI-CVQKD protocol, aiming to eliminate the negative effect of AMs in the MDI-CVQKD system. 

As shown in Fig. \ref{MDI}(c), the PP DPM-based MDI-CVQKD protocol can be designed from the PP MDI-CVQKD protocol by replacing AM to an extra fiber link with a PM and a FM at Alice's and Bob's sides. This architecture totally removes AMs so that one does not need to consider its practical problem in the experimental implementation. In what follows, we show that the PP DPM-based MDI-CVQKD protocol is equivalent to the Gaussian-modulated MDI-CVQKD protocol. 

Generally, the Jones matrix of a Faraday mirror can be expressed by
\begin{equation}
\textbf{J}_{FM}=
 \left[
 \begin{matrix}
   \mathrm{cos}\, \theta & \mathrm{sin} \,\theta \\
   -\mathrm{sin}\,\theta & \mathrm{cos}\,\theta  \\
  \end{matrix}
  \right]
   \left[
 \begin{matrix}
   1 & 0 \\
   0 & -1  \\
  \end{matrix}
  \right]
   \left[
 \begin{matrix}
   \mathrm{cos}\, \theta & -\mathrm{sin} \,\theta \\
   \mathrm{sin}\,\theta & \mathrm{cos}\,\theta  \\
  \end{matrix}
  \right] =    
  \left[
 \begin{matrix}
   \mathrm{cos}\, (2\theta) & -\mathrm{sin} \,(2\theta) \\
   -\mathrm{sin}\,(2\theta) & -\mathrm{cos}\,(2\theta)  \\
  \end{matrix}
  \right].
\end{equation}
When the input signal reaches FM and reflects back \cite{ref42}, the complete Jones matrix of the rotated element can be expressed by 
\begin{equation}
\textbf{R}=\textbf{T}(-\delta)\textbf{J}_{FM}\textbf{T}(\delta)=e^{i(\varphi_{o}+\varphi_{e})}\textbf{J}_{FM},
\end{equation}
where $\delta$ is the rotation angle between the reference basis and the eigenmode basis of the birefringence medium, while $\varphi_{o}$ and $\varphi_{e}$ are the propagation phases of ordinary and extraordinary rays, respectively. $\textbf{T}(\pm\delta)$ are the Jones matrices of birefringence medium when the signal goes forward and backward of the single-mode delay lines, which can be given by 
\begin{equation}
\textbf{T}(\pm\delta)=
 \left[
 \begin{matrix}
   \mathrm{cos}\, \delta & \mp\mathrm{sin} \,\delta \\
   \pm\mathrm{sin}\,\delta & \mathrm{cos}\, \delta  \\
  \end{matrix}
  \right]
   \left[
 \begin{matrix}
   e^{i\varphi_o} & 0 \\
   0 & e^{i\varphi_e}  \\
  \end{matrix}
  \right]
   \left[
 \begin{matrix}
   \mathrm{cos}\, \delta & \pm\mathrm{sin} \,\delta \\
   \mp\mathrm{sin}\,\delta & \mathrm{cos}\, \delta  \\
  \end{matrix}
  \right].
\end{equation}
Since the PP DPM-based MDI-CVQKD protocol can be constructively symmetric, we consider Alice's data-processing for simplicity. The transformation matrices of the dual-phase modulation scheme can be given by \cite{ref32}
\begin{equation}
\textbf{J}_{PM_{A1}+FM_{A1}}=\textbf{T}(-\delta)\textbf{J}_{PM_{A1x}}\textbf{R}\textbf{J}_{PM_{A1y}}\textbf{T}(\delta)=\varsigma_{A1}e^{i(\varphi_{A1})}\textbf{R},
\end{equation}
\begin{equation}
\textbf{J}_{PM_{A2}+FM_{A2}}=\textbf{T}(-\delta)\textbf{J}_{PM_{A2x}}\textbf{R}\textbf{J}_{PM_{A2y}}\textbf{T}(\delta)=\varsigma_{A2}e^{i(\varphi_{A2})}\textbf{R}, 
\end{equation}
where $\varsigma_{A1}$ and $\varsigma_{A2}$ are the equivalent attenuation coefficient of PM$_{A1}$ and PM$_{A2}$, $\varphi_{A1}$ and $\varphi_{A2}$ are electronically modulated phases of PM$_{A1}$ and PM$_{A2}$. Suppose the input Jones vector is $Alice_{in}$, the output of dual-phase modulation $Alice_{out}$ after a round trip can be expressed as
\begin{equation}
\begin{aligned}
Alice_{out}&=\frac{1}{2}Alice_{in}(\textbf{J}_{PM_{A1}+FM_{A1}}+\textbf{J}_{PM_{A2}+FM_{A2}}) \\
&=\frac{1}{2}(\varsigma_{A1}Alice_{in}e^{i(\varphi_{A1})}+\varsigma_{A2}Alice_{in}e^{i(\varphi_{A2})})\textbf{R}.
\end{aligned}
\end{equation}
In an ideal dual-phase modulation system with perfect optical components, one can get the same insertion loss in the two arms. Namely we have $\varsigma\approx\varsigma_{A1}\approx\varsigma_{A2}$, and then the output of dual-phase modulation can be simplify as
\begin{equation}
\begin{aligned}
Alice_{out}&=\varsigma Alice_{in}\mathrm{exp}\left[\frac{i(\varphi_{A1}+\varphi_{A2})}{2}\right]\mathrm{cos}\left(\frac{\varphi_{A1}-\varphi_{A2}}{2}\right)\textbf{R}.
\end{aligned}
\label{Alice}
\end{equation}
Similarly, the output of dual-phase modulation at Bob's side can be given by
\begin{equation}
\begin{aligned}
Bob_{out}&=\varsigma Bob_{in}\mathrm{exp}\left[\frac{i(\varphi_{B1}+\varphi_{B2})}{2}\right]\mathrm{cos}\left(\frac{\varphi_{B1}-\varphi_{B2}}{2}\right)\textbf{R}.
\end{aligned}
\label{Bob}
\end{equation}
According to Eq.(\ref{Alice}) and Eq.(\ref{Bob}), we find that the Gaussian modulation can be implemented by both senders using two polarization-independent PMs instead of a polarization-dependent AM and PM. Therefore, the PP DPM-based  MDI-CVQKD protocol is equivalent to the Gaussian-modulated MDI-CVQKD protocol, leading to the convenient experimental implementation in efficiency.  

\section{Security Analysis}

In this section, we analyze the security of the PP DPM-based MDI-CVQKD  protocol in both asymptotic case \cite{ref43} and finite-size regime\cite{ref44,ref45}. We find that almost all raw keys generated by the proposed scheme can be used for the final secret key generation when considering the finite-size effect. 

\subsection{Asymptotic security of PP DPM-based MDI-CVQKD protocol}
\begin{figure}[htbp]
\centering\includegraphics[width=13cm]{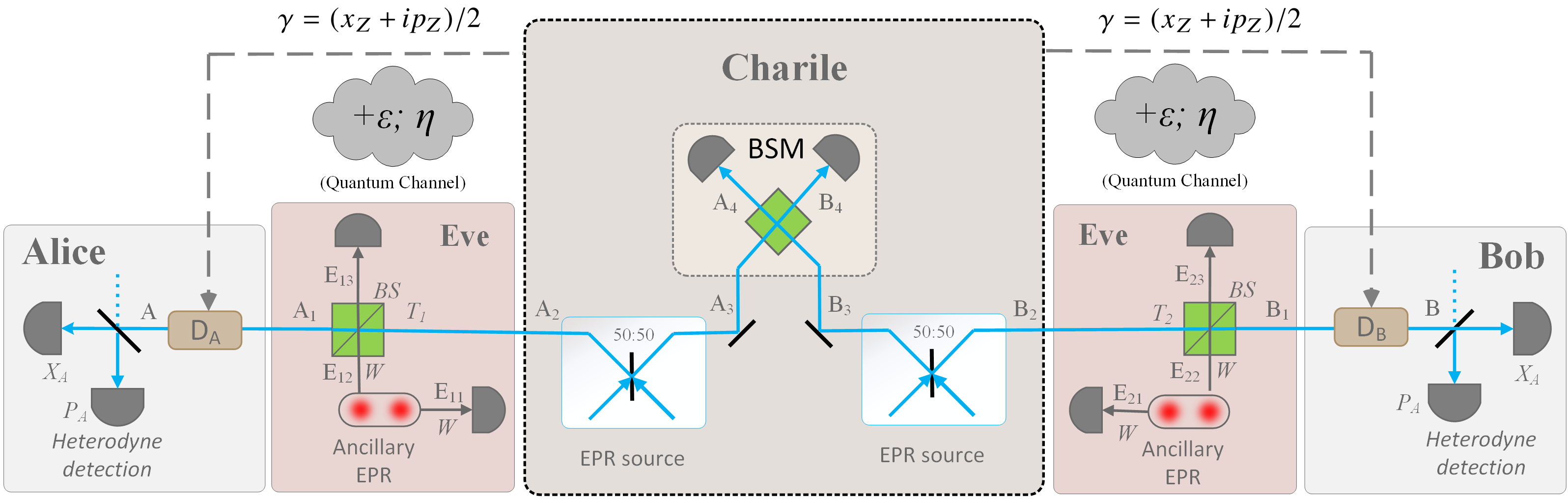}
\caption{Entanglement-based model of PP DPM-based MDI-CVQKD protocol with entangling cloner attacks. Charlie prepares two EPR pairs and sends one mode of each to Alice and Bob through the untrusted quantum channel, respectively. Alice and Bob displace the incoming modes according to the public BSM result and subsequently measure them with respective heterodyne detections.}
\label{MDIsec2}
\end{figure}

As shown in Fig.(\ref{MDIsec2}), we depicts the entanglement-based model of the PP DPM-based MDI-CVQKD protocol. Since only one laser is used for preparing coherent states at Charlie's side instead of adopting two separate lasers at each remote side, the source can be modeled by using two EPR pairs at Charlie's side. After Alice and Bob respectively displace their incoming modes ($A_1$ and $B_1$) according to the BSM results that are publicly announced by Charlie, mode $A_2$ and mode $B_2$ have the certain correlation. Providing that mode $A_2$ and mode $B_2$ come from the same EPR pair, it is similar to the CVQKD protocol with an entangled source in the middle \cite{ref38,ref46}.

We assume that Eve performs the collective Gaussian attack strategy, which is the best attack under the direct and reverse reconciliation protocols \cite{ref36,ref37}. In particular, Eve prepares her ancillary system in a product state for Alice's and Bob's side, and the ancilla mode of each side interacts individually with a single pulse sent to Alice and Bob respectively. The combined state reads
\begin{equation}
\begin{aligned}
&\rho_{A_2E_{1}B_2E_{2}} =
 \sum_{a,b}[P(a)|a\rangle\langle a|\otimes\psi_{A_2E_{1}}^{a} \oplus P(b)|b\rangle\langle b|\otimes\psi_{B_2E_{2}}^{b}]^{\otimes n}.
\end{aligned}
\end{equation}
Eve then launches the entangling cloner attack. Specifically, Eve replaces the channel with transmittance $T$ and excess noise referred to the input $\chi$ by preparing the ancilla $|E_{i}\rangle$ of variance $W_{i}$ ($i=1,2$) and a beam splitter of transmittance $T$. The value $W_{i}$ can be tuned to match the noise of the real channel $\chi=(1-T)/T+\varepsilon$. Note that for the PP DPM-based MDI-CVQKD protocol, both sides of Alice and Bob are symmetric, i.e., $T=T_{1}=T_{2}$. After that, Eve keeps one mode $E_{i1}$ of $|E_{i}\rangle$ and injects the other mode $E_{i2}$ into the unused port of each beam splitter and thus acquires the output mode $E_{i3}$. After repeating this process for each pulse, Eve stores her ancilla modes, $E_{i1}$ and $E_{i3}$, in quantum memories. Finally, Eve measures the exact quadrature on $E_{i1}$ and $E_{i3}$ after Charlie reveals the BSM results. 

According to the above-mentioned situation, the lower bound of the asymptotic secret key rate under the collective attack strategy can be given by
\begin{equation}
\begin{aligned}
K_{asym}=\beta I(A:B)-\chi_{E},
\end{aligned}
\end{equation}
where $\beta$ is the reconciliation efficiency, $I(A:B)$ is the Shannon mutual information between Alice and Bob, and $\chi_{E}$ is the Holevo bound of Eve's information  \cite{ref47}. Assuming that Alice and Bob have perfect heterodyne detectors, the covariance matrix of the Gaussian state $\rho_{AB}^{G}$ can be given by
\begin{equation}       
\Gamma_{AB}^{G}\!\!=\!\!
\left[                 
  \begin{array}{cc}   
    a\mathbb{I} & c\sigma_{z}\\  
    c\sigma_{z} & b\mathbb{I}\\  
  \end{array}
\right]=
\left[                 
  \begin{array}{cc}   
    [T_{1}V+(1-T_{1})W_{1}]\mathbb{I} & \sqrt{T_{1}T_{2}(V^{2}-1)}\sigma_{z}\\  
    \sqrt{T_{1}T_{2}(V^{2}-1)}\sigma_{z} & [T_{2}V+(1-T_{2})W_{2}]\mathbb{I}\\  
  \end{array}
\right],                 
\end{equation}
where $\mathbb{I}$ and $\sigma_{z}$ represent $\mathrm{diag}(1,1)$ and $\mathrm{diag}(1,-1)$, respectively, $V$ is the variance of mode $A$ and mode $B$, $W_{i}=T_{i}\chi_{i}/(1-T_{i})$ with the added noise referred to the input $\chi_{i}=(1-T_{i})/T_{i}+\varepsilon$. Therefore, Alice and Bob's mutual information can be calculated as
\begin{equation}
\begin{aligned}
I(A:B)=\mathrm{log}_{2}\left[\frac{b+1}{b+1-c^{2}/(a+1)}\right]
\end{aligned}.
\end{equation}

As the proposed protocol is symmetric, the orientation of reconciliation would not have affect on the performance of the protocol, and thus we only consider the calculation of asymptotic security key rate in direct reconciliation (the identical result can be obtained in reverse reconciliation). The mutual information between Alice and Eve can be expressed as
\begin{equation}
\begin{aligned}
\chi_E=S(E)-S(E|A)
\end{aligned}.
\end{equation}
Due to the fact that Eve can provide a purification of Alice and Bob's density matrix, we obtain $S(E)=S(AB)$, which is a function of the symplectic eigenvalues $\lambda_{1,2}$ of $\Gamma_{AB}^{G}$ given by
\begin{equation}
\begin{aligned}
S(AB)=G[(\lambda_{1}-1)/2]+G[(\lambda_{2}-1)/2]
\end{aligned},
\end{equation}
where
$G(x)=(x+1)\mathrm{log}_{2}(x+1)-x\mathrm{log}_{2}x$ is the Von Neumann entropy and the symplectic eigenvalues $\lambda_{1,2}$ are calculated as
\begin{equation}
\begin{aligned}
\lambda_{1,2}^{2}=\frac{1}{2}[\Delta\pm\sqrt{\Delta^{2}-4D^{2}}]
\end{aligned},
\end{equation}
with $\Delta=a^{2}+b^{2}-2c^{2}$ and $D=ab-c^{2}$. 
After Alice performs heterodyne detection over mode $A$, the system $BE$ is pure. This gives $S(E|A)=S(B|A)=G[(\lambda_{3}-1)/2]$, where the symplectic eigenvalue $\lambda_3=b-c^2/(a+1)$. See \cite{ref48} for the detailed derivations.

\begin{figure}[htbp]
\centering\includegraphics[width=9cm]{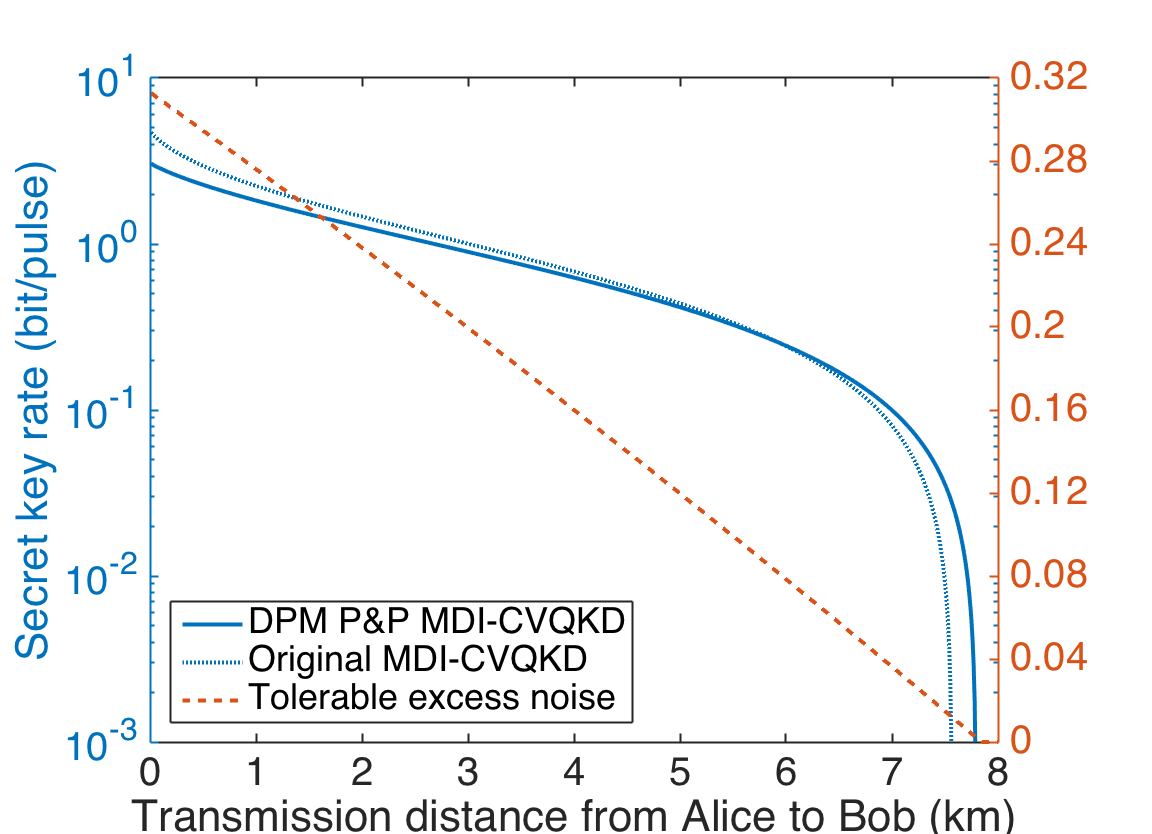}
\caption{The performance of MDI-CVQKD protocols. Blue solid line and red dashed line denote the asymptotic secret key rate and the tolerable excess noise of the proposed PP DPM-based MDI-CVQKD protocol as a function of transmission distance from Alice to Bob, respectively. As a comparison, blue dotted line denotes the asymptotic secret key rate of traditional MDI-CVQKD protocol in \cite{ref27}. The simulation parameters are set as follows: modulation variance is $V=20$, reconciliation efficiency is $\beta=95\%$ and excess noise for blue solid line is $\varepsilon=0.001$.}
\label{MDITEN}
\end{figure}

Fig. (\ref{MDITEN}) shows the performance of the PP DPM-based MDI-CVQKD protocol in asymptotic case. As a comparison, we also plot the asymptotic secret key rate of the traditional MDI-CVQKD protocol \cite{ref27}. We find that the performances of both protocols are similar except for a few minor discrepancies, i.e. the slight differences in maximal secret key rate and maximal transmission distance. It shows that the PP DPM-based MDI-CVQKD protocol features the same security level as the traditional MDI-CVQKD protocol. It is worth noticing that the aim of this scheme is not to improve the performance of the MDI-CVQKD protocol, but to show the feasibility and substitutability of the MDI-CVQKD protocol in experimental implementation. As a result, we only consider the symmetric case of MDI-CVQKD protocol, i.e., $L_{AC}=L_{BC}$,  regardless of the asymmetric one, although the latter case can largely improve the transmission distance of the MDI-CVQKD protocol \cite{ref26,ref27}.

\subsection{Security of the PP DPM-based MDI-CVQKD protocol in finite-size regime}

The above asymptotic security proof is based on an assumption that one considers the security of a protocol in asymptotic limit of infinitely many signals that are transmitted between Alice and Bob.  However, it is unpractical for implementations. Fortunately, a security framework of finite-size for the CVQKD protocols has been proposed in \cite{ref49}. In this framework, the raw key is no longer infinite and one needs to use part of it to estimate the parameters of the communication channel. However, it would introduce a tradeoff between the final secret key rate and the accuracy of parameter estimation step in the finite-size regime. Very recently, \cite{ref50} shows that this problem can be solved in traditional MDI-CVQKD protocol. We, here, extend it to the proposed PP DPM-based MDI-CVQKD protocol and give the detailed security proof in finite-size regime.

An important procedure of data post-processing is parameter estimation, aiming to acquire the information of quantum channel such as transmissivity and excess noise, which are relevant for estimating the security of the CVQKD protocol. In general, local information without classical communication is not sufficient for parameter estimation, and the only way for obtaining the precise result of parameter estimation is to sacrifice part of raw key. However, the more raw key data are used for parameter estimation, the lower is the final secret key rate. In fact, the estimation of the covariance matrix $\Gamma_{XYZ}$ of the tripartite state $\rho_{XYZ}$ could be done locally by Alice or Bob without using part of the raw key. In particular, the covariance matrix $\Gamma_{XYZ}$ can be expressed as
\begin{equation}       
\Gamma_{XYZ}=
\left[                 
  \begin{array}{cccccc}   
    \textbf{X} & \textbf{0} & \textbf{c}_{XZ} \\  
    \textbf{0} & \textbf{Y} & \textbf{c}_{YZ}\\  
    \textbf{c}_{XZ}^{\intercal} & \textbf{c}_{YZ}^{\intercal} & \textbf{Z}\\ 
  \end{array}
\right],                 
\end{equation}
where 
$     
\textbf{X}=\textbf{Y}=
\left[                 
  \begin{array}{cc}   
     V & 0 \\
     0 & V \\
  \end{array}
\right]$,  the matrix
\begin{equation}    
\textbf{Z}=
\left[                 
  \begin{array}{cc}   
     \langle x_Z^2\rangle & \langle x_Zp_Z\rangle \\
     \langle x_Zp_Z\rangle & \langle p_Z^2 \rangle\\
  \end{array}
\right],                 
\end{equation}
is the empirical covariance matrix of $(x_Z, p_Z)$, and 
\begin{equation}       
\textbf{c}_{XZ}=
\left[                 
  \begin{array}{cc}   
    \langle x_A'x_Z\rangle & \langle x_A'p_Z\rangle \\  
    \langle p_A'x_Z\rangle & \langle p_A'p_Z\rangle \\  
  \end{array}
\right],       \quad            
\textbf{c}_{YZ}=
\left[                 
  \begin{array}{cc}   
    \langle x_B'x_Z\rangle & \langle x_B'p_Z\rangle\\
    \langle p_B'x_Z\rangle & \langle p_B'p_Z\rangle \\
  \end{array}
\right]             
\end{equation}
are the correlation items.

Since the proposed protocol is based on the MDI-CVQKD structure, where Alice and Bob modulate coherent states using the DPM scheme and do not perform any measurement at their own side, the variances of $x_A'$, $p_A'$, $x_B'$ and $p_B'$ can be known locally by Alice and Bob. After Charlie announces the measurement result $\gamma=(x_Z+ip_Z)/2$, Alice computes the empirical correlations of the matrix $\textbf{c}_{XZ}$, namely $\langle x_A'x_Z\rangle$, $\langle x_A'p_Z\rangle$, $\langle p_A'x_Z\rangle$ and $\langle p_A'p_Z\rangle$. Similarly, Bob can obtain the empirical correlations of the matrix $\textbf{c}_{YZ}$. As a result, all the entries of the covariance matrix $\Gamma_{XYZ}$ can be calculated locally by Alice and Bob without any extra communication. Finally, the covariance matrix $\Gamma_{AB}^G$ of the Gaussian state $\rho_{AB}^G$ can be achieved by exploiting the relations Eq. (1). Note that the amplification coefficient $k$ has to be well selected to optimalize the conditional displacements in both Alice's and Bob's sides \cite{ref27,ref50}. 

Based on the derived covariance matrix $\Gamma_{AB}^G$, the performance of the PP DPM-based MDI-CVQKD protocol can be estimated by Alice or Bob in finite-size regime. Specifically, the secret key rate calculated by taking finite-size effect into account is expressed as \cite{ref49}
\begin{equation}\label{fini1}
\begin{aligned}
K_{fini}=\frac{n}{N}[\beta I(A:B)-S_{\epsilon_{PE}}-\Delta(n)],
\end{aligned}
\end{equation}
where $\beta$ and $I(A:B)$ are as the same as the afore-mentioned definitions, $\epsilon_{PE}$ is the failure probability of parameter estimation and the parameter $\Delta(n)$ is related to the security of the privacy amplification given by
\begin{equation}
\begin{aligned}
\Delta(n)=(2\mathrm{dim}\mathcal{H}+3)\sqrt{\frac{\mathrm{log}_2(2/\bar{\epsilon})}{n}}+\frac{2}{n}\mathrm{log}_2(1/\epsilon_{PA}),
\end{aligned}
\end{equation}
where $\bar{\epsilon}$ is a smoothing parameter, $\epsilon_{PA}$ is the failure probability of privacy amplification, and $\mathcal{H}$ is the Hilbert space corresponding to the raw key. Since the raw key is usually encoded on binary bits, we have $\mathrm{dim}\mathcal{H}=2$. We denote $N$ the total exchanged signals and $n$ the number of signals that is used for sharing key between Alice and Bob. Note that in the conventional calculation the remained  $m=N-n$ signals are used for parameter estimation so that the values are usually set to be $m=n=\frac{1}{2}N$. However, as we pointed above, the remained signals can be neglected since parameter estimation can be locally performed without extra information. Therefore, the signals that are used for estimating can be exploited for transporting more secret keys. That is to say, almost all raw keys can be used for the final secret key generation without parameter estimation using part of them, leading to the increased secret key rate of the MDI-CVQKD protocol. In fact, Alice and Bob still need to share the entries of the estimated covariance matrix, which contain an amount of raw keys. Fortunately, that amount is negligible as the secret key is very long. As a result, the value can be set to $n\approx N$.

In the conventional finite-size case (needing to sacrifice part of raw keys), $S_{\epsilon_{PE}}$ needs to be calculated in parameter estimation procedure where one can find a covariance matrix $\Gamma_{\epsilon_{PE}}$ minimizing the secret key rate with a probability of $1-\epsilon_{PE}$. It can be calculated by $m$ couples of correlated variables $(x_i,y_i)_{i=1\cdots m}$  given by
\begin{equation}       
\Gamma_{\epsilon_{PE}}\!\!=\!\! 
\left(                 
 \begin{array}{cc}   
   V\mathbb{I} & tZ\sigma_{z}\\  
    tZ\sigma_{z} & (t^2 V+\sigma^2)\mathbb{I}\\  
  \end{array}
\right),         
\end{equation}
where $t=\sqrt{\eta}$ and $\sigma^2=1+\eta\varepsilon$ are compatible with $m$ sampled data except with probability $\epsilon_{PE}/2$. The maximum-likelihood estimators $\hat{t}$ and $\hat{\sigma^2}$ respectively has the follow distributions
\begin{equation}
\begin{aligned}
\hat{t}\sim \big(t,\; \frac{\sigma^2}{\sum_{i=1}^mx_i^2}\big)\quad \mathrm{and} \quad \frac{m\hat{\sigma}^2}{\sigma^2}\sim\chi^2(m-1),
\end{aligned}
\end{equation}
where $t$ and $\sigma^2$ are the authentic values of the parameters. In order to maximize the value of the Holevo information obtained by Eve with the statistics except with probability $\epsilon_{PE}$, the value of $t_{min}$ (the lower bound of $t$) and $\sigma_{max}^2$ (the upper bound of $\sigma^2$) in the limit of $m$ must be computed, namely
\begin{equation}
\begin{aligned}
t_{min}&=\sqrt{\eta}-z_{\epsilon_{PE}/2}\sqrt{\frac{1+\eta\varepsilon}{mX'}}, \\
\sigma_{max}^2&=1+\eta\varepsilon+z_{\epsilon_{PE}/2}\frac{\sqrt{2}(1+\eta\varepsilon)}{\sqrt{m}},
\end{aligned}
\end{equation}
where $z_{\epsilon_{PE}/2}$ is such that $1-\mathrm{erf}(z_{\epsilon_{PE}/2}/\sqrt{2})/2=\epsilon_{PE}/2$ and erf is the error function defined as
\begin{equation}
\begin{aligned}
\mathrm{erf}(x)=\frac{2}{\pi}\int_0^xe^{-t^2}\mathrm{d}t.
\end{aligned}
\end{equation}
The above-mentioned error probabilities can be set to
$
\bar{\epsilon}=\epsilon_{PE}=\epsilon_{PA}=10^{10}.
$
Finally, one can derive the secret key rate using the derived bounds $t_{min}$ and $\sigma_{max}^2$.

Actually, we do not need to estimate the secret key rate of the PP DPM-based MDI-CVQKD protocol like above, but would directly calculate it by the locally obtained covariance matrix $\Gamma_{XYZ}$ without complicated estimation process. This is feasible since the correlations between Alice's and Bob's raw keys are post-selected by the relay so that the public variable $Z$ contains all the information about the correlations between Alice and Bob \cite{ref50}.

\begin{figure}[htbp]
\centering\includegraphics[width=10cm]{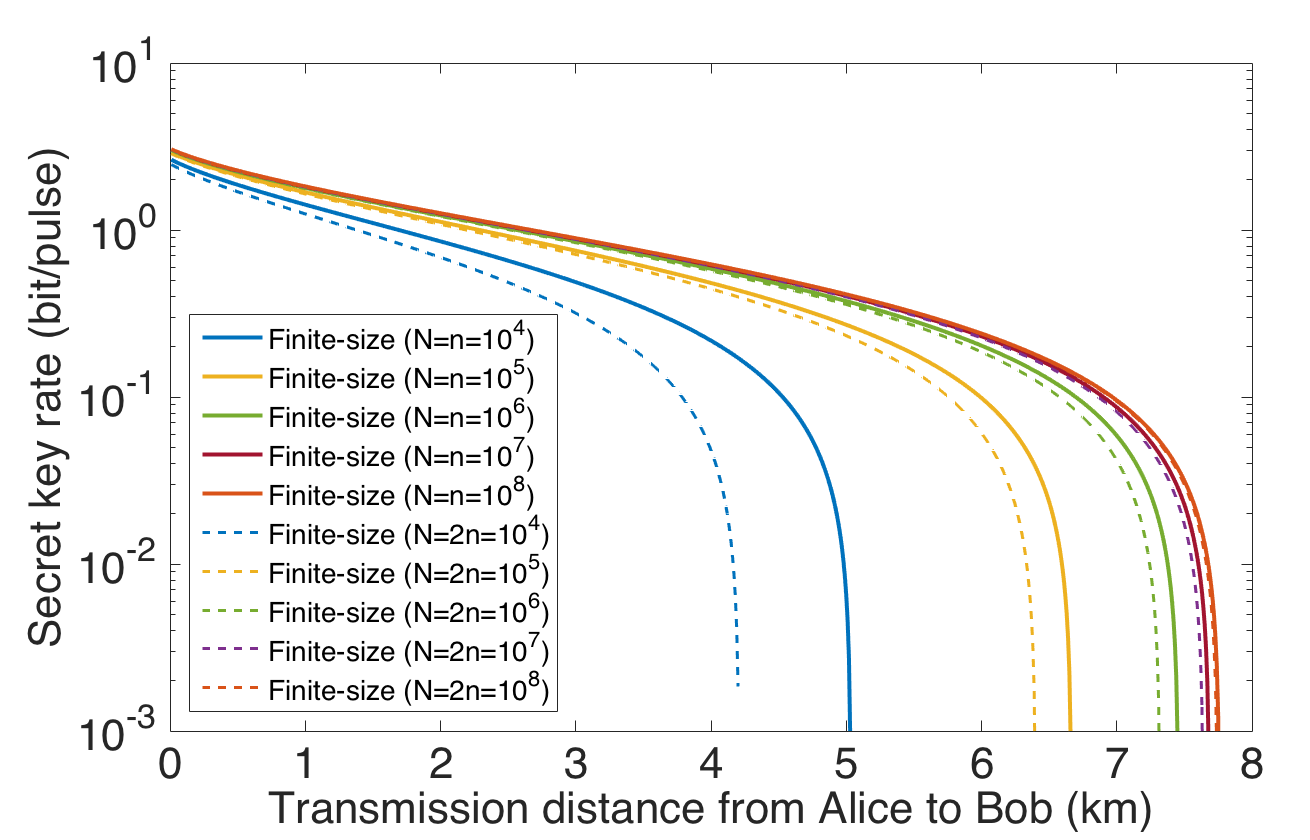}
\caption{Finite-size secret key rate of the proposed PP DPM-based MDI-CVQKD protocol as a function of transmission distance. Solid lines show the secret key rate generated from almost all raw keys ($N=n$), dashed lines show the secret key rate using conventional finite-size calculation ($N=2n$). From left to right, both lines correspond to block lengths of $N=10^4$, $10^5$, $10^6$, $10^7$ and $10^8$. The parameters are set as same as Fig. \ref{MDITEN}.}
\label{finite}
\end{figure}

Fig. \ref{finite} shows the performance of the PP DPM-based MDI-CVQKD protocol with almost all raw keys are used for generating the final secret key (solid lines) comparing with conventional finite-size calculation (dashed lines). We find that for each block, especially for the small-length block, the maximal transmission distance can be extended by directly calculating the locally obtained covariance matrix, since part of raw key data that should be used for parameter estimation now is used for generating more final secret keys, giving birth to the increased secret key rate of the MDI-CVQKD protocol using conventional finite-size calculation.

We note that the CVQKD protocols has been recently proved to be secure against collective attacks in a composable security framework \cite{ref51}, which is the enhancement of security based on uncertainty of the finite-size effect so that one can obtain the tightest secure bound of the protocol by considering each data-processing step in the CVQKD system \cite{ref52}. We do not give the detailed proof of the composable security for MDI-based CVQKD protocols here, but it is reasonable to believe that the performance of the proposed protocol can be improved either since the conventional proof of composable security also needs to sacrifice part of raw keys for parameter estimation.

\section{Discussion and conclusion}

\begin{figure}[htbp]
\centering
\includegraphics[width=12cm]{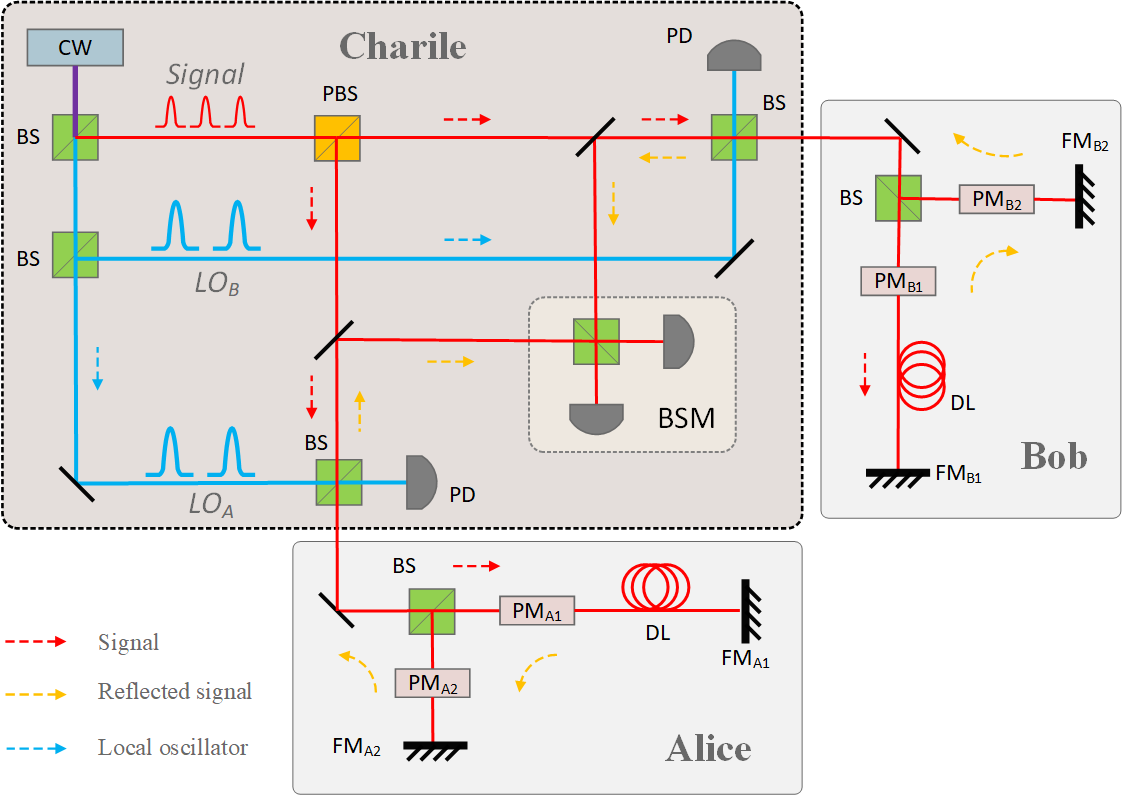}
\caption{Experimental concept of the PP DPM-based MDI-CVQKD protocol. CW, Continuous-wave laser; BS, Beam splitter; PM, Phase modulator; PBS, Polarizing beam splitter; DL, Delay line; FM, Faraday mirror; PD, Photoelectric detector; BSM, Bell state measurement.}
\label{exp}
\end{figure}

So far we have illustrate the characteristics of the PP DPM-based MDI-CVQKD protocol. As for its practical implementations, we demonstrate its setup shown in Fig. \ref{exp}. Charlie generates a series of strong pulses using continuous-wave (CW) laser. These pulses are splitted into two portions with an intensity ratio of 99:1, a fraction (1\%, red line) of which is used to carry signals while another fraction (99\%, blue line) is used as a locally generated LO. The signals are divided into two branches by the PBS and then sent to Alice and Bob respectively. For both Alice and Bob, after being reflected by FM$_1$, the signals are modulated by the DPM scheme and reversely sent back to Charlie. The incoming modulated signals, subsequently, are interfered with respective local LOs, aiming to calibrate the incoming signals and monitor its variance in real time. Finally, the yielded signals from Alice and Bob are used for the BSM.

There are several remarks on the proposed protocol for its implementation. First of all, all LO-aimed attacks e.g. wavelength attacks, saturation attacks, calibration attacks and LO fluctuation attacks, can be well defended due to fact that the LO is locally generated by the trusted part Charlie. Secondly, the synchronization problem of Alice and Bob is eliminated because both signal and LO come from the same laser. Thirdly, the reference of two signals can be guaranteed identically and the polarization drifts can be compensated automatically since only one laser is required for the proposed scheme. Moreover, there is no need to use the traditional LiNbO$_3$ modulators by applying DPM scheme, which takes advantage of the polarization-insensitive properties of phase modulators so that the coherent-state preparation would not be affected by the polarization drifts of the fiber channel.

In conclusion, we have suggested the design of the PP DPM-based  MDI-CVQKD with no extra performance penalty. The proposed scheme waives the necessity of propagation of LO through the insecure quantum channels. Because a real local LO can be generated from the same laser of quantum signal at Charlie's side, it avoids the problems of synchronization of different lasers as well as the LO-aimed attacks. Moreover, the reference of two signals can be guaranteed identically and the polarization drifts can be compensated automatically since only one laser is required for the proposed scheme. Meanwhile, a polarization-insensitive dual-phase modulation strategy is adopted to Alice's and Bob's sides respectively, which shows the experimental feasibility of preparing Gaussian states. We  derive the security bounds against optimal Gaussian collective attacks. It shows that the proposed scheme works equivalently to symmetrically modulated Gaussian-state MDI-CVQKD protocols. Since almost entire raw keys generated by the proposed scheme can be used for final secret key generation when considering the finite-size effect, the secret key rates of MDI-CVQKD in finite-size regime can be increased. Moreover, an experimental concept of the proposed scheme, which can be deemed guideline for final implementation, is demonstrated. In terms of possible future research, we will give the concrete experimental implementation of the PP DPM-based MDI-CVQKD protocol.

\section*{Funding}
This work is supported by the National Natural Science Foundation of China (Grant No. 61379153, No. 61572529).

\section*{Acknowledgments}
We would like to thank Professor S. Pirandola for his helpful suggestion.

\section*{Disclosures}
The authors declare that there are no conflicts of interest related to this article.

\end{document}